\documentclass[12pt]{article}

\usepackage[a4paper,left=30mm,right=20mm,top=20mm,bottom=20mm]{geometry}
\usepackage{graphics}
\usepackage{amsmath}
\usepackage{epsfig}
\usepackage{cite}
\usepackage{xcolor}
\usepackage[unicode]{hyperref}

\title{Particle Event Generator: A Simple-in-Use System PEGASUS version 1.0 }

\author{A.V.~Lipatov$^{1,2}$, M.A.~Malyshev$^{1}$, S.P.~Baranov$^{3}$}

\begin{document}

\maketitle

\begin{center}

{\it $^{1}$Skobeltsyn Institute of Nuclear Physics, Lomonosov Moscow State University, 119991, Moscow, Russia}\\
{\it $^{2}$Joint Institute for Nuclear Research, 141980, Dubna, Moscow region, Russia}\\
{\it $^{3}$P.N.~Lebedev Physics Institute, 119991 Moscow, Russia}\\

\end{center}

\vspace{0.5cm}

\begin{center}

{\bf Abstract }

\end{center}

\indent
\textsc{pegasus} is a parton-level Monte-Carlo event generator
designed to calculate cross sections for a wide range of hard QCD processes
at high energy $pp$ and $p\bar p$ collisions, which incorporates the dynamics of transverse 
momentum dependent (TMD)
parton distributions in a proton. Being supplemented with off-shell production
amplitudes for a number of partonic subprocesses
and provided with necessary TMD gluon density functions, it produces weighted or unweighted event records
which can be saved as a plain data file or a file
in a commonly used Les Houches Event format.
A distinctive feature of the \textsc{pegasus} is an intuitive and 
extremely user friendly interface, allowing one to easily implement various kinematical cuts
into the calculations.
Results can be also presented "on the fly" with built-in tool \textsc{pegasus plotter}.
A short theoretical basis is presented and 
detailed program description is given.

\vspace{1cm}


\newpage

\section{Program summary} \indent

\noindent
{\it Title:} \textsc{pegasus 1.0}

\vspace{0.3cm}

\noindent
{\it Computer for which the program is designed and others on which it is operable:} Linux systems

\vspace{0.3cm}

\noindent
{\it Programming Language used:} C++ and Fortran 77, compiled with g++ and gfortran

\vspace{0.3cm}

\noindent
{\it High-speed storage required:} No

\vspace{0.3cm}

\noindent
{\it Separate documentation available:} No

\vspace{0.3cm}

\noindent
{\it Keywords:} QCD, BFKL and CCFM evolution equations, small-$x$ physics, TMD parton densities, high-energy factorization

\vspace{0.3cm}

\noindent
{\it Physical problem:}
Theoretical description of a number of high energy processes proceeding with large momentum 
transfer and containing multiple hard scales needs for 
transverse momentum dependent (TMD) parton (quark
or gluon) distributions in a proton\cite{1,2}. 
These quantities encode the nonperturbative information on proton structure, 
including transverse momentum and polarization degrees of freedom and
satisfy the Balitsky-Fadin-Kuraev-Lipatov (BFKL)\cite{3} or 
Catani-Ciafaloni-Fiorani-Marchesini (CCFM)\cite{4} evolution
equations, which resum large logarithmic terms proportional to
$\alpha_s^n \ln^n s/\Lambda_{\rm QCD}^2 \sim \alpha_s^n \ln^n 1/x$.
At high energies (or, equivalently, at small $x$) the latter 
are expected to become equally (or even more) important
than the conventional Dokshitzer-Gribov-Lipatov-Altarelli-Parisi (DGLAP)
contributions proportional to $\alpha_s^n \ln^n \mu^2/\Lambda_{\rm QCD}^2$\cite{5}.
The CCFM equation takes into account additional terms proportional to
$\alpha_s^n \ln^n 1/(1 - x)$ and, therefore, is valid at 
both small and large $x$.

\vspace{0.3cm}

\noindent
{\it Solution:}
Experimental investigations of hadron scattering processes at 
high energy colliders, such as the Large Hadron Collider (LHC), usually
imply multiparticle final states and implement complex kinematical 
restrictions on the fiducial phase space. 
The multi-purpose Monte-Carlo event generators
are commonly used tools for 
theoretical description of the collider measurements.
Most of them (for example, \textsc{pythia 8.2}\cite{6}, \textsc{mcfm 9.0}\cite{7}, 
\textsc{madgraph5}\_a\textsc{mc@nlo}\cite{8}
and other) use conventional (collinear) QCD factorization, which is 
based on the DGLAP resummation.
On the other side, the hadron-level Monte-Carlo event generator \textsc{cascade}\cite{9} and
parton-level generator \textsc{katie}\cite{10} can deal with non-zero
transverse momenta of incoming off-shell partons. In particular, \textsc{cascade} employs
the CCFM equation for initial state gluon evolution.
\textsc{pegasus} is a newly developed parton-level Monte-Carlo event generator
designed to calculate cross sections for a wide range of hard QCD processes, 
which also incorporates the TMD gluon dynamics in a proton.
It provides all necessary components, including 
off-shell (dependent on the transverse momenta) production
amplitudes and grid files for TMD gluon
densities, interpolated automatically.
\textsc{pegasus} offers an intuitive and 
extremely user friendly interface, which allows one 
to easily implement various kinematical cuts into the calculations.
Generated events (weighted or unweighted) can be stored in the 
Les Houches Event file or presented "on the fly" with convenient
built-in tool \textsc{pegasus plotter}.

\vspace{0.3cm}

\noindent
{\it Restrictions on the complexity of the physical problem:}
The following production amplitudes are implemented:
\begin{itemize}
\item $g^* + g^* \to Q + \bar Q$, where $Q = c$ or $b$
\item $g^* + g^* \to {\cal Q} + b + \bar c$, where ${\cal Q} = B_c$ or $B_c^{(*)}$
\item $g^* + g^* \to Q\bar Q\left[\,^3S_1^{(1)}\right] + g \to {\cal Q} + g$, where ${\cal Q} = \psi^\prime$, $J/\psi$ or $\Upsilon(3S)$
\item $g^* + g^* \to Q\bar Q\left[\,^1S_0^{(8)}, \,^3S_1^{(8)}, \, ^3P_J^{(8)}\right] \to {\cal Q}$, where ${\cal Q} = \psi^\prime$, $J/\psi$ or $\Upsilon(3S)$
\item $g^* + g^* \to Q\bar Q\left[\,^3P_J^{(1)}, \,^3S_1^{(8)} \right] \to {\cal Q}$, where ${\cal Q} = \chi_{cJ}(1P)$ or $\chi_{bJ}(3P)$
\item $g^* + g^* \to H^0 \to \gamma \gamma$
\item $g^* + g^* \to H^0 \to ZZ^* \to 4l$
\item $g^* + g^* \to H^0 \to W^+W^- \to e^{\pm}\mu^{\mp} \nu \bar \nu$
\item $g^* + g^* \to V + Q + \bar Q$, where $V = \gamma$ or $V = Z/\gamma^*$
\item $q + g \to V + q$, where $V = \gamma$ or $V = Z/\gamma^*$
\item $q + Q \to V + q + Q$, where $V = \gamma$ or $V = Z/\gamma^*$
\item $q + \bar q \to V + Q + \bar Q$, where $V = \gamma$ or $V = Z/\gamma^*$
\end{itemize}

\noindent
The standard spectroscopic notation $^{(2S+1)}L_J^{(a)}$ is used for 
intermediate Fock state of heavy quark pair $Q\bar Q$ produced 
with spin $S$, orbital angular momentum $L$, total angular momentum $J$
and color representation $a$.
In the subprocesses above, the radiative decays 
$\chi_{cJ}(1P) \to J/\psi + \gamma$ and $\chi_{bJ}(3P) \to \Upsilon(3S) + \gamma$
are simulated keeping the proper spin structure of the decay amplitudes.
The radiative decays $\psi^\prime \to J/\psi + \pi^+ + \pi^-$ or 
$\psi^\prime \to J/\psi + \pi^0 + \pi^0$ can be generated according to the phase space. 
Also, the subsequent leptonic decays 
$J/\psi \to l^+l^-$, $\Upsilon(3S) \to l^+l^-$ and $\psi^\prime \to l^+l^-$, $Z/\gamma^* \to l^+l^-$ 
can be generated optionally.

The present version of \textsc{pegasus} is applicable for Fermilab Tevatron and CERN LHC processes.
\vspace{0.3cm}

\noindent
{\it Other program used:}
\textsc{qwtplot} library (version 6.1.3)\cite{11} for histogram plotting, 
\textsc{vegas} routine\cite{12} for multi-dimensional Monte-Carlo 
integration, stand-alone C++ codes from 
MMHT\cite{13} and CTEQ\cite{14} groups
to provide access to the MSTW'2008, MMHT'2014, CT'14 and CTEQ6.6 
parton density functions (implemented into the program package).

\vspace{0.3cm}

\noindent
{\it Download of the program:} \url{https://theory.sinp.msu.ru/dokuwiki/doku.php/pegasus/download}

\section{Theoretical framework} \indent

Here we present a short review of ideas
and frameworks which form the theoretical basis of \textsc{pegasus}.
For more information we address the reader to reviews\cite{1,2}. 

\subsection{Kinematics and cross section formula} \indent

\textsc{pegasus} operates in the general framework of conventional Parton Model extended to $k_T$-factorization 
approach\cite{15,16}. It employs the factorization hypothesis to calculate the cross section
of a physical process through the convolution of the (TMD or collinear) 
parton densities and hard scattering amplitudes. By default, 
the $k_T$-factorization scheme is assumed
and can be switched to the collinear QCD factorization
for each of colliding particles.

\textsc{pegasus} refers to $2\to 1$, $2\to 2$ and $2\to 3$ partonic subprocesses.
The initial partonic state is described with Sudakov variables, namely, 
the light-cone momentum fractions $x_1$ and $x_2$ and non-zero parton transverse
momenta $k_{1T}$ and $k_{2T}$. The $n$-particle final state phase space is
parameterized in terms of particle rapidities $y_i$, transverse momenta 
$p_{iT}$, and azimutal angles $\phi_i$, where $i = 1\,...\,n$\cite{17}.
The fully differential cross section reads 
\begin{equation}
  \displaystyle d\sigma(p p \to p_1 \, ... \, p_n + X) = {\pi \over \hat s F} (4\pi)^{2 - 2n} \sum_{a, b} \sum_{\rm spin} \sum_{\rm colors} |{\cal A}_{ab}(k_1 k_2 \to p_1 \, ... \,p_n)|^2 \times \atop {
  \displaystyle \times f_a(x_1, {\mathbf k}_{1T}^2,\mu^2) f_b(x_2, {\mathbf k}_{2T}^2,\mu^2) d{\mathbf k}_{1T}^2 d{\mathbf k}_{2T}^2 d{\mathbf p}_{1T}^2 \, ... \, d{\mathbf p}_{(n-1)T}^2 dy_1 \, ... \, dy_n {d\phi_1 \over 2\pi} \, ... \, {d\phi_{n-1} \over 2\pi} },
\end{equation}
\noindent
where the 4-momenta for all particles are given in parentheses, $\hat{s}=(k_1+k_2)^2$ is the subprocess invariant energy, 
$F$ is the flux factor (see below) and $\mu^2$ is the factorization scale.
The longitudinal momentum fractions $x_1$ and $x_2$ are not the
integration variables; they are obtained from the energy-momentum conservation laws in the light-cone projections:
\begin{equation}
  x_1 \sqrt{s} = \sum_{i = 1}^n m_{iT} \exp(y_{i}), \quad x_2 \sqrt{s} = \sum_{i = 1}^n m_{iT} \exp(-y_{i}),
\end{equation}
\noindent
where $m_{iT}$ is the transverse mass of produced particle $i$.
In the optional choice of collinear QCD factorization, we replace the 
TMD parton densities with conventional distributions, omit the integration over ${\mathbf k}_{1T}^2$ 
and ${\mathbf k}_{2T}^2$ and take the on-shell limit in the 
scattering amplitude ${\cal A}_{ab}$ as described below.

\subsection{Off-shell partonic amplitudes} \indent

The calculation of partonic amplitudes follows the standard Feynman rules,
with the exception that the initial gluons are off-shell. 

Off-shell gluons may have nonzero transverse momentum and an admixture 
of longitudinal component in the polarization vector. In accordance with 
the $k_T$-factorization prescriptions, the initial gluon spin density matrix 
is taken in the form\cite{16}:
\begin{equation} \label{gauge}
  \sum \epsilon_g^\mu \epsilon_g^{*\nu} = k_T^\mu k_T^\nu/|k_T|^2.
\end{equation}
\noindent
In the collinear limit, when $k_T\to 0$, this expression converges to the ordinary 
$\sum \epsilon_g^\mu \epsilon_g^{*\nu} = -g^{\mu\nu}/2$. 
This property provides continuous on-shell limit for the partonic
amplitudes.

In general, off-shell initial states may cause violation of gauge
invariance in the non-abelian theories. We solve this problem in a way\cite{18}. 
We start with a set of "extended" diagrams, where the off-shell gluons
are considered as emitted by on-shell external fields (say, quarks), 
so that they represent internal lines in the Feynman graphs. As all
of the external lines in these graphs are on shell, the gauge invariance 
of the whole set is fulfilled. However, the extended gauge invariant set
may contain unfactorizable diagrams that cannot be represented as a
convolution of the hard scattering amplitude and gluon density functions 
(such as, for example, Figs. 5(e) and 5(d) in\cite{18}). 
At the same time, we learn from\cite{18} that the non-factorizable 
diagrams vanish in the particular gauge (\ref{gauge}).
Therefore, within this gauge, we are left with the usual diagrams of the
Parton Model.

In fact, the prescription (\ref{gauge}) is a remake of Equivalent Photon
Approximation (EPA) in QED and DIS. Consider a photon emitted by an electron: 
$e(p)\to e'(p')+ \gamma(k)$. 
Then, taking trace over the electron line in the matrix element squared 
one obtains the polarization tensor
\begin{equation}
 L^{\mu\nu}= {\rm tr} \left[(\hat p'+m_e)\,\gamma^\mu\,(\hat p+m_e)\,\gamma^\nu\right] = 8\,p^\mu p^\nu - 4(pk)\,g^{\mu\nu}.
\end{equation}
\noindent
Neglecting the second term in the right hand side in the small $x$ limit,
$p\gg k$, one immediately arrives at the spin structure
$\sum \epsilon^{\mu}\epsilon^{*\nu} \sim L^{\mu\nu}\sim p^\mu p^\nu$.
It can be rewritten in the form (\ref{gauge}) after using the
Sudakov representation $k=xp+k_{T}$ and applying a gauge shift 
$\epsilon^\mu\to\epsilon^\mu-k^\mu/x$.
The gauge invariance of the matrix element is correct to the
accuracy of ${\cal O}(x)\simeq 10^{-4} - 10^{-6}$.

\subsection{CCFM evolution equation} \indent

The CCFM evolution equation\cite{4} for TMD gluon density
resums both BFKL\cite{3} large logarithms $\alpha_s^n \ln^n 1/x$ and
$\alpha_s^n \ln^n 1/(1 - x)$ terms.
In the limit of asymptotic energies,
it is almost equivalent to BFKL, but also similar to usual
DGLAP evolution\cite{5} for large $x$.
Moreover, it introduces angular
ordering condition to treat correctly gluon
coherence effects. In the leading logarthmic approximation (LLA), the CCFM equation
for TMD gluon density $f_g(x,{\mathbf k}_T^2,\mu^2)$
can be written as
\begin{equation}
  \displaystyle f_g(x,{\mathbf k}_T^2,\mu^2) = f_g^{(0)}(x,{\mathbf k}_T^2,\mu_0^2) \Delta_s(\mu,\mu_0) + \atop {
  \displaystyle + \int\frac{dz}{z}\int\frac{dq^2}{q^2}\Theta(\mu-zq)\Delta_s(\mu,zq) \tilde P_{gg}(z,{\mathbf k}_T^2, q^2) f_g\left(\frac{x}{z},{\mathbf k}^{\prime \, 2}_T,q^2\right) },
\end{equation}

\noindent
where ${\mathbf k}_T^\prime = {\mathbf q}(1 - z) + {\mathbf k}_T$
and $\tilde P_{gg}(z,{\mathbf k}^2_T,q^2)$ is the CCFM
splitting function:
\begin{equation}
  \displaystyle \tilde P_{gg}(z,{\mathbf k}^2_T,q^2) = \bar\alpha_s(q^2(1-z)^2) \left[\frac{1}{1-z}+\frac{z(1-z)}{2}\right] + \atop {
  \displaystyle + \bar\alpha_s({\mathbf k}_T^2)\left[\frac{1}{z}-1+\frac{z(1-z)}{2}\right]\Delta_{ns}(z,{\mathbf k}^2_T,q^2) }.
\end{equation}

\noindent
The Sudakov and non-Sudakov form factors read:
\begin{equation}
 \ln \Delta_s(\mu,\mu_0)= - \int\limits_{\mu_0^2}^{\mu^2}\frac{d\mu^{\prime \, 2}}{\mu^{\prime \, 2}}\int\limits_0^{1-\mu_0/\mu^{\prime}}d\zeta\,\frac{\bar\alpha_s(\mu^{\prime \, 2}(1-\zeta)^2)}{1-\zeta},
\end{equation}
\begin{equation}
\ln \Delta_{ns}(z,{\mathbf k}_T^2, {\mathbf q}_T^2) = -\bar\alpha_s({\mathbf k}_T^2)\int\limits_0^1\frac{dz^\prime}{z^\prime}\int\frac{dq^2}{q^2}\Theta({\mathbf k}_T^2-q^2)\Theta(q^2-z^{\prime\,2} {\mathbf q}^2_T).
\end{equation}

\noindent
where $\bar \alpha_s = 3 \alpha_s/\pi$.
The first term in the CCFM equation is the initial
TMD gluon density $f_g^{(0)}(x,{\mathbf k}_T^2,\mu_0^2)$ multiplied by the Sudakov form factor,
decribing the contribution of non-resolvable branchings between
the starting scale $\mu_0^2$ and scale $\mu^2$.
The second term represents the details of the QCD evolution
expressed by the convolution of the CCFM gluon splitting function $\tilde P_{gg}(z,{\mathbf k}^2_T,q^2)$
with the gluon density and Sudakov form factor. The angular ordering condition
is introduced with the theta function, so
the evolution scale $\mu^2$ is defined by the maximum allowed
angle for any gluon emission\cite{4}.
The main advantage of this approach is the implicit
including of higher-order radiative corrections (namely, a part
of NLO + NNLO +... terms corresponding to the initial state real gluon
emissions).
Some details can be found,
for example, in reviews\cite{2}.

A similar equation can be also written\cite{19} for TMD valence quark
densities with replacement of the gluon splitting function by the
quark one\footnote{The TMD sea quark densities are not defined in
CCFM approach. However, they can be obtained from the gluon ones in
the last gluon splitting approximation\cite{20}.}.
One usually takes the initial TMD gluon and valence quark distributions as
\begin{equation}
\label{fitA}
  f_g^{(0)}(x, {\mathbf k}_T^2, \mu_0^2) = Nx^{-B}(1-x)^C\exp(-{\mathbf k}_T^2/\sigma^2),
\end{equation}
\begin{equation}
  f_{q_v}^{(0)}(x, {\mathbf k}_T^2, \mu_0^2) = x q_v(x, \mu_0^2) \exp(-{\mathbf k}_T^2/\sigma^2)/\sigma^2,
\end{equation}

\noindent
where $\sigma=\mu_0/\sqrt{2}$ and $q_{v}(x,\mu^2)$ is the
conventional (collinear) quark density function.
The parameters $N$, $B$ and $C$ can be fitted from collider data (see\cite{19}).

The TMD (valence) quark densities for CCFM approach are not available
in the current version of the \textsc{pegasus}.

\subsection{Parton Branching approach} \indent

The Parton Branching (PB) method\cite{21,22} allows one to solve the DGLAP equations iteratively.
It gives a possibility to take into account simultaneously soft-gluon
emission at $z \to 1$ and transverse momentum $\mathbf {q}_T$
recoils in the parton branchings along the QCD cascade.
The latter leads to a natural determination
of TMD density functions for both gluons and quarks.
A soft-gluon resolution scale $z_M$ is introduced
to separate resolvable and non-resolvable emissions,
which are treated via the DGLAP splitting functions $P_{ab}(\alpha_s, z)$
and Sudakov form-factors, respectively.
The PB equations for TMD parton densities read:
\begin{equation}
  \displaystyle f_a(x,{\mathbf k}_T^2,\mu^2) = f_a^{(0)}(x,{\mathbf k}_T^2,\mu_0^2)\Delta_a(z_M,\mu^2,\mu^2_0) + \sum_b\int\limits_x^{z_M}dz\int\frac{d^2\mathbf q}{\pi\mathbf q^2}\Theta(\mu^2- \mathbf q^2) \times \atop{
  \displaystyle \times \Theta(\mathbf q^2-\mu_0^2) \frac{\Delta_a(z_M,\mu^2,\mu^2_0)}{\Delta_a(z_M,\mathbf q^2,\mu^2_0)} P_{ab}^{(R)}(\alpha_s(\mathbf q^2), z) f_b\left(\frac{x}{z},{\mathbf k}_T^{\prime \,2},q^2\right) },
\end{equation}

\noindent
where $a = q$ or $g$ and ${\mathbf k}_T^\prime = {\mathbf q}(1 - z) + {\mathbf k}_T$.
The Sudakov form factors are defined as
\begin{equation}
  \ln \Delta_a(z_M,\mu^2,\mu^2_0)= -\sum_b\int\limits_{\mu_0^2}^{\mu^2}\frac{d\mu^{\prime \, 2}}{\mu^{\prime \,2}}\int\limits_0^{z_M}d\zeta\,\zeta\,P_{ba}^{(R)}(\alpha_s(\mu^{\prime \, 2}),\zeta).
\end{equation}

\noindent
The real-emission branching probabilities $P_{ab}^{(R)}$ are obtained from splitting functions $P_{ab}$
by eliminating $\delta(1-z)$-terms and substituting $1/(1-z)_+\to1/(1-z)$.
The evolution scale $\mu^2$ can be connected with the emission
angle with respect to the beam direction, that
leads to the angular ordering condition $\mu = |{\mathbf q}_T|/(1 - z)$.
The dependence on the $z_M$
dissapears when this angular ordering condition is applied and $z_M$ is
large enough.
The initial TMD parton distributions are taken in a factorized form
as a product of conventional quark and gluon densities and
intrinsic transverse momentum distributions (taken in gaussian
form~\cite{8,9}), where all the parameters can be fitted from the collider data.
Unlike the CCFM, the PB densities have the strong normalization property:
\begin{equation}
  \int\limits_0^{\mu^2} f_a(x, {\mathbf k}_T^2,\mu^2) d{\mathbf k}_T^2 = xa(x,\mu^2),
\end{equation}

\noindent
where $a = q$ or $g$ and $xa(x,\mu^2)$ are the 
conventional parton density functions (PDFs).
The PB equations can be solved numerically by an iterative Monte-Carlo
method with the leading order (LO) or next-to-leading order (NLO) accuracy. 
The solution results in a steep decline of the parton
densities at ${\mathbf k}_T^2 > \mu^2$.
It contrasts the CCFM evolution,
where the transverse momentum is allowed to be larger
than the scale $\mu^2$, corresponding to an effective taking into account 
higher-order contributions\footnote{Very recently,
a method to incorporate CCFM effects into the PB
formulation has been proposed\cite{23}.}.

\subsection{Kimber-Martin-Ryskin approach} \indent

The Kimber-Martin-Ryskin (KMR) approach\cite{24} provides a technique to construct TMD gluon and quark
densities from conventional ones by loosing the DGLAP strong ordering
condition at the last evolution step, that results in $k_T$ dependence of the parton distributions.
This procedure is believed to take into account effectively
the major part of next-to-leading logarithmic (NLL) terms $\alpha_s (\alpha_s \ln\mu^2)^{n-1}$
compared to the LLA, where
terms proportional to $\alpha_s^n \ln^n \mu^2$ are taken into account.

At the LO, the KMR method, defined for ${\mathbf k}_T^2 \geq \mu_0^2 \sim 1$~GeV$^2$, results in expressions for TMD 
quark and gluon distributions\cite{24}:
\begin{gather}
 \displaystyle f_q(x, {\mathbf k}_T^2,\mu^2) = T_q({\mathbf k}_T^2,\mu^2)\frac{\alpha_s({\mathbf k}_T^2)}{2\pi}\times \atop{
 \displaystyle \int\limits_x^1dz\left[P_{qq}^{\rm LO}(z)\frac{x}{z}q\left(\frac{x}{z},{\mathbf k}_T^2\right)\Theta\left(\Delta-z\right)+P_{qg}^{\rm LO}(z)\frac{x}{z}g\left(\frac{x}{z},{\mathbf k}_T^2\right)\right] },\\
 \displaystyle f_g(x, {\mathbf k}_T^2,\mu^2) = T_g({\mathbf k}_T^2,\mu^2)\frac{\alpha_s({\mathbf k}_T^2)}{2\pi}\times \atop{
  \displaystyle \int\limits_x^1dz\left[\sum_qP_{gq}^{\rm LO}(z)\frac{x}{z}q\left(\frac{x}{z},{\mathbf k}_T^2\right)+P_{gg}^{\rm LO}(z)\frac{x}{z}g\left(\frac{x}{z},{\bf k}_T^2\right)\Theta\left(\Delta-z\right)\right] },
\end{gather}
\noindent
where $P_{ab}^{\rm LO}(z)$ are the usual DGLAP splitting functions at LO
and $\mu_0^2$ is the minimum scale for which DGLAP evolution is valid.
The theta functions introduce the specific ordering conditions in the last evolution step, 
thus regulating the soft gluon singularities.
The cut-off parameter $\Delta$ usually has one of two forms, $\Delta = \mu/(\mu + |{\mathbf k}_T|)$ or 
$\Delta = |{\mathbf k}_T|/\mu$, that reflects the angular or strong ordering conditions, respectively.
In the case of angular ordering, the parton densities are extended into the
${\mathbf k}_T^2 > \mu^2$ region, whereas the strong ordering condition leads to a steep drop
of the parton distributions beyond the scale $\mu^2$.
At low ${\mathbf k}_T^2 < \mu_0^2$ the behaviour of the TMD parton densities has 
to be modelled\cite{24}.
Usually it is assumed to be flat under strong normalization condition~(9).

The Sudakov form factors allow one to include logarithmic virtual (loop) corrections, they take the form:
\begin{gather}
T_q({\mathbf k}_T^2,\mu^2)=\exp\left[-\int\limits_{{\mathbf k}_T^2}^{\mu^2}\frac{d{\mathbf q}_T^2}{{\mathbf q}_T^2}\frac{\alpha_s({\mathbf q}_T^2)}{2\pi}\int\limits_0^{z_\text{max}}d\zeta\,P_{qq}^\text{LO}(\zeta)\right],\\
T_g({\mathbf k}_T^2,\mu^2)=\exp\left[-\int\limits_{{\mathbf k}_T^2}^{\mu^2}\frac{d{\mathbf q}_T^2}{{\mathbf q}_T^2}\frac{\alpha_s({\mathbf q}_T^2)}{2\pi}\left(\int\limits_{z_\text{min}}^{z_\text{max}}d\zeta\,\zeta P_{gg}^\text{LO}(\zeta)+n_f\int\limits_0^1d\zeta\,P_{gq}^\text{LO}(\zeta)\right)\right],
\end{gather}
with $z_\text{max}=1-z_\text{min}=\mu/(\mu+|{\mathbf q}_T|)$. These form factors
give the probability of evolving from a scale ${\mathbf k}_T^2$ to a scale $\mu^2$ without parton emission.
At the NLO, the TMD parton densities can be written as\cite{25}:
\begin{equation}
 f_a(x,{\mathbf k}_T^2,\mu^2)=\int\limits_0^1dz\,T_a({\mathbf p}_T^2,\mu^2)\frac{\alpha_s({\mathbf p}_T^2)}{2\pi} \sum_{b=q,g}P_{ab}^\text{NLO}(z)\frac{x}{z}b\left(\frac{x}{z},{\mathbf p}_T^2\right)\Theta(\Delta - z),
\end{equation}
where ${\mathbf p}_T^2={\mathbf k}_T^2/(1-z)$. Note that both DGLAP splitting functions and 
conventional parton distributions should be taken with NLO accuracy. The Sudakov form factors at NLO read:
\begin{gather}
T_q({\mathbf k}_T^2,\mu^2)=\exp\left[-\int\limits_{{\mathbf k}_T^2}^{\mu^2}\frac{d{\mathbf q}_T^2}{{\mathbf q}_T^2}\frac{\alpha_s({\mathbf q}_T^2)}{2\pi}\int\limits_0^1d\zeta\,\zeta (P_{qq}^\text{NLO}(\zeta)+P_{gq}^\text{NLO}(\zeta))\right],\\
T_g({\mathbf k}_T^2,\mu^2)=\exp\left[-\int\limits_{{\mathbf k}_T^2}^{\mu^2}\frac{d{\mathbf q}_T^2}{{\mathbf q}_T^2}\frac{\alpha_s({\mathbf q}_T^2)}{2\pi}\int\limits_0^1d\zeta\,\zeta (P_{gg}^\text{NLO}(\zeta)+2n_fP_{qg}^\text{NLO}(\zeta))\right].
\end{gather}

\noindent
However, it was demonstrated that the NLO prescription, with a good accuracy, can be significantly simplified to keep only 
the LO splitting functions\cite{25} while the main effect is related to the Sudakov form factors.

\subsection{Flux factor} \indent

The choice of the flux factor is another peculiarity of off-shell
calculations.
The definition of the flux, which is the velocity of an off-shell particle, 
is highly disputable and is not clear. By default, we accepted the analytic
continuation of the general textbook definition\cite{17}:
\begin{equation}\label{flux}
  F = 2 \lambda^{1/2}(\hat{s}, k_1^2, k_2^2).
\end{equation}
\noindent
Our choice is supported by a numerical experiment\cite{26}, 
where we have considered the production of $\chi_{cJ}$ states in a two
photon process, $e + e\to\chi_{cJ} + e'{+}e'$ and made a comparison between
the prompt calculation of this $2\to 3$ process and calculation based 
on EPA, $\gamma + \gamma\to\chi_{cJ}$ with $J = 0$, $1$ or $2$. 
We find 
that the flux taken in the form (\ref{flux})
provides a sensibly better fit to the exact calcualtion. The fact that 
the exact calculation disagrees with the factorized (collinear) 
calculation indicates that the conditions for the factorization theorem 
are yet not fulfilled. 
In such a situation, our choice of the flux can be regarded as a
phenomenological correction for non-factorizable contributions. The same
definition of the flux is adopted, for example, in\cite{10}.
However, the user can optionally choose the conventional 
definition $F = 2 x_1 x_2 s$, as it is explained below.

\section{Calculations using PEGASUS} \indent

\textsc{pegasus} has an intuitive and unprecedentedly
user friendly interface.
The calcualtions using \textsc{pegasus} include
a few general steps common for all of the processes.
So, when \textsc{pegasus} is running, one can select 
the colliding particles, $pp$ or $p\bar p$, and set their 
center-of-mass energy $\sqrt s$.
The default setting corresponds to the LHC Run II setup.
Then one can select factorization scheme (TMD or collinear one)
for each of the colliding particles, choose corresponding
parton density function and set the parameters, important 
for further Monte-Carlo simulation, namely,
number of iterations and number of simulated events per iteration (see Fig.~1).
Next steps could be as follows (note that all these steps do not 
depend on each other and can be done in different order):
\begin{itemize}
\item From the list of available processes one can select 
the necessary process and then (optionally) correct the default kinematical 
restrictions, hard scales, list of requested observables and corresponding binnings (see Figs.~2 --- 4).
This can be done by double clicking on the requested process or via main 
menu (using {\sl Edit $\to$ Task} option).
\item For each of the observables one 
can manually edit the default binning according to own wishes. 
As another option, the binning can be uploaded immediately from the data file.
Several commonly used formats (such as \texttt{.yoda}, \texttt{.yaml}, \texttt{.csv} 
or plain data format, compatible with \textsc{gnuplot}\cite{27} tool)
as provided by HepData repository\cite{28} are supported.
\item The user-defined setup for any process 
(total center-of-mass energy, selected parton densities, kinematical restrictions, binnings etc)
can be saved to a configuration file in some internal format (\texttt{*.pegasus}).
This can be done via the main menu (using {\sl File $\to$ Save} or {\sl File $\to$ Save As} options) or
via the popup menu available on right mouse button click or via appropriate button in the button panel.
Of course, the configuration file can be loaded and a user-defined setup can 
be used in further applications.
This can be done via main menu (with {\sl File $\to$ Open} option) or via 
popup menu or via {\sl Open} button on the button panel. 
\item Weighted or unweighted events can be generated. This option
is available via main menu {\sl Edit $\to$ Settings $\to$ Generated events}
or via popup menu.
\item If one needs to generate the Les Houches Event file\cite{29},
one has to mark corresponding option before 
the calculation starts (see Fig.~1).
Note that this option affects the speed of the calculations.
\item The calculation will start by choosing the corresponding option in 
main menu ({\sl Calculation $\to$ Start}), popup menu or pressing {\sl Start} button on the button panel.
The numerical results for requested observables will be immediately presented
"on the fly" with built-in tool \textsc{pegasus plotter} (see Fig.~5).
The calculations can be paused or even stopped (using main menu options 
{\sl Calculation $\to$ Pause}, {\sl Calculation $\to$ Stop}, corresponding buttons 
on the button panel or options in popup menu).
Of course, during pause in the calculations, all manipulations with accumulated 
results in \textsc{pegasus plotter} are allowed (see Section~4.3).
\item If there are several contributing subprocesses, there is a possibility 
to immediately jump to the next one (via {\sl Calculation $\to$ Next} option in main menu
or appropriate button on button panel or popup menu) during the calculations.
\end{itemize}

\noindent
The generated events can be accumulated in Les Houches Event (\texttt{*.lhe})
file and/or presented in \textsc{pegasus plotter}.
Using the latter, one can save the results in a some internal format (\texttt{*.pplot})
or as a simple plain data (compatible, for example, with \textsc{gnuplot} package)
or export them to an image (\texttt{*.pdf}, \texttt{*.png}, \texttt{*.jpg} or \texttt{*.bmp}).

Below we give a more detailed information and
explanations about the important features of \textsc{pegasus}.

\subsection{Parton density functions in a proton} \indent

Latest sets\cite{19} of CCFM-evolved TMD gluon distributions in a proton,
JH'2013 set 1 and JH'2013 set 2, are available
in the \textsc{pegasus}.
The input parameters of JH'2013 set 1 
were determined from the fit to high precision HERA data on the proton
structure function $F_2(x,Q^2)$, whereas the parameters of JH'2013 set 2 
gluon were extracted from combined fit on both $F_2(x,Q^2)$ and 
$F_2^c(x,Q^2)$ data (see \cite{19}).
The previous CCFM fits, namely, A0 and B0 sets\cite{30}, are available also
and comparison between them can be found\cite{19}.
Technically, all these CCFM-evolved gluon densities are stored on a grid
in $\log x$, $\log k_T$ and $\log \mu$ and a simple linear
interpolation is applied to evaluated the gluon density for values
in between the grid points. This interpolation
proceeds automatically at the each event generation.
The data files containing the grid points are
supplied with the program package and 
read in at the beginning of each requested calculation\footnote{We would like to note that, in 
the case of CCFM equation, the TMD gluon distribution can be derived from a forward evolution 
procedure as implemented in the \textsc{updfevolv}
routine\cite{31}. From 
the initial gluon density as given by (9), which includes a Gaussian 
intrinsic $k_T$ distribution, a set 
of values $x$ and $k_T$ can be obtained by evolving
up to a given scale $\mu$. 
The input parameters in (9) have to be fitted from
the data.}.

The TMD gluon and quark densities can be also evaluated 
in the standard DGLAP scenario using the PB and/or KMR schemes (with LO or NLO accuracy).
As an input for the KMR procedure, the conventional (collinear)
PDFs have to be applied.
Several sets of KMR and PB-based gluon densities are currently available in \textsc{pegasus}.
So, well known NNPDF3.1 (LO)\cite{32} and MMHT'2014 (LO)\cite{13} parametrizations 
and recent analytical expressions obtained\cite{33,34,35} 
in the so-called generalized double asymptotic scaling (DAS)
approximation of QCD\cite{34,35} were used as an input.
The DAS approximation is connected to the asymptotic behaviour
of the DGLAP evolution discovered many years ago\cite{36}. 
Flat initial conditions for the DGLAP equations, applied
in the generalized DAS scheme, lead to the Bessel-like behaviour for the
proton PDFs at small $x$\cite{34,35}.
The DAS LO set 1 corresponds to "frozen" treatment
of the QCD strong coupling in the infrared region: 
$\alpha_s(\mu^2) \to \alpha_s(\mu^2 + m_\rho^2)$.
The DAS LO set 2 is based on the idea\cite{37} regarding 
the  analyticity of the strong coupling at low scales.
The difference between these two choices is discussed\cite{38}.
Everywhere, the cut-off parameter $\Delta$ is taken according to 
angular ordering condition.

\begin{table}
\footnotesize
\label{table1}
\begin{center}
\begin{tabular}{|l|c|c|c|c|}
\hline
 & & & &\\
  Set & Order of $\alpha_s$ & $N_f$ & $\Lambda_{\rm QCD}$/MeV & Ref. \\
 & & & &\\
\hline
 & & & &\\
  A0 (CCFM) & 1 & 4 & 250  & \cite{30} \\
 & & & &\\
  B0 (CCFM) & 1 & 4 & 250  & \cite{30} \\
 & & & &\\
  JH'2013 set 1 (CCFM) & 2 & 4 & 200  & \cite{19} \\
 & & & &\\
  JH'2013 set 2 (CCFM) & 2 & 4 & 200  & \cite{19} \\
 & & & &\\
  KMR (MMHT'2014 LO) & 1 & 5 & 211  & \cite{24} \\
 & & & &\\
  KMR (NNPDF3.1 LO) & 1 & 5 & 167  & \cite{24} \\
 & & & &\\
  KMR (DAS LO set 1) & 1 & 4 & 143  & \cite{38} \\
 & & & &\\
  KMR (DAS LO set 2) & 1 & 4 & 143  & \cite{38} \\
 & & & &\\
  PB-NLO-HERAI+II'2018 set 1 & 2 & 4 & 118  & \cite{22} \\
 & & & &\\
  PB-NLO-HERAI+II'2018 set 2 & 2 & 4 & 118  & \cite{22} \\
 & & & &\\
\hline
\end{tabular}
\end{center}
\caption{The TMD gluon densities in a proton implemented into the \textsc{pegasus}. The 
A0$\pm$, B0$\pm$, JH'2013 set 1$\pm$ and JH'2013 set 2$\pm$ distributions, needed
to estimate the scale uncertainties of the calculations (see below), are not shown.}
\end{table}

The available TMD gluon sets with the essential
parameters are listed in Table~1.
We note that large variety of the TMD gluon 
distribution functions in a proton are collected in the \textsc{tmdlib}
package\cite{39}, which is a C++ library providing a framework
and an interface to the different parametrizations. 

To perform the calculations in the collinear
QCD factorization (mainly at LO or tree-level NLO for some processes)
several sets of conventional PDFs are available in \textsc{pegasus}.
These are recent MMHT'2014 (LO and NLO)\cite{13} and CT14 (LO and NLO)\cite{40} ones
and previous sets provided by CTEQ Collaboration (CTEQ'6.6)\cite{41},
MSTW'2008 (LO and NLO)\cite{42} and GRV'94 (LO)\cite{43}.
The standalone C++ codes from MMHT\cite{13} and CTEQ\cite{14} groups
are implemented into the \textsc{pegasus} code
and corresponding data files containing the grid points
are supplied with the program package.

\subsection{Simulated processes} \indent

\begin{table}
\footnotesize
\label{table1}
\begin{center}
\begin{tabular}{|l|c|c|c|c|}
\hline
 & & & &\\
  Process  & Subprocesses  & Scales & Observables & Ref. \\
 & & & &\\
\hline
 & & & &\\
  Open heavy flavor production  & $g^* + g^*\to Q + \bar Q$ & $p_T(Q)$ & $p_T(Q)$              & \cite{44} \\
  ~\textbullet~charm            &                           & $m_T(Q)$ & $y(Q)$                & \\
  ~\textbullet~beauty           &                           &          & $M(Q\bar Q)$          & \\
                                &                           &          & $\Delta\phi(Q\bar Q)$ & \\
 & & & &\\
\hline
 & & & &\\
  Double flavored bound state   & $g^* + g^* \to {\cal Q} + b + \bar c$ & $p_T({\cal Q})$ & $p_T({\cal Q})$ & \cite{45} \\
  production                    &                                       & $m_T({\cal Q})$ & $y({\cal Q})$   & \\
  ~\textbullet~$B_c$            &                                       &                 &                 & \\
  ~\textbullet~$B_c^{(*)}$      &                                       &                 &                 & \\
 & & & &\\
\hline
 & & & &\\
  $S$-wave heavy quarkonia      & $g^* + g^* \to Q\bar Q[^3S^{(1)}_1] + g \to {\cal Q} + g$ & $p_T({\cal Q})$ & $p_T({\cal Q})$ & \cite{46, 47} \\
  production                    & $g^* + g^* \to Q\bar Q[^1S^{(8)}_0] \to {\cal Q}$         & $m_T({\cal Q})$ & $y({\cal Q})$   & \\
  ~\textbullet~$J/\psi$         & $g^* + g^* \to Q\bar Q[^3S^{(8)}_1] \to {\cal Q}$         &                 &                 & \\
  ~\textbullet~$\psi^\prime$    & $g^* + g^* \to Q\bar Q[^3P^{(8)}_J] \to {\cal Q}$         &                 &                 & \\
  ~\textbullet~$\Upsilon(3S)$   &                                          &                 &                 & \\
 & & & &\\
\hline
 & & & &\\
  $P$-wave heavy quarkonia      & $g^* + g^* \to Q\bar Q[^3P^{(1)}_J] \to {\cal Q}$ & $p_T({\cal Q})$ & $p_T({\cal Q})$ & \cite{46, 47} \\
  production                    & $g^* + g^* \to Q\bar Q[^3S^{(8)}_1] \to {\cal Q}$ & $m_T({\cal Q})$ & $y({\cal Q})$   & \\
  ~\textbullet~$\chi_{cJ} (1P)$ &                                                   &                 &                 & \\
  ~\textbullet~$\chi_{bJ} (3P)$ &                                                   &                 &                 & \\
 & & & &\\
\hline
 & & & &\\
 Inclusive Higgs production                                                & $g^* + g^* \to H^0$ & $m(H^0)$   & $p_T(H^0)$ & \cite{48} \\
  ~\textbullet~$H^0 \to \gamma \gamma$                            &                     & $m_T(H^0)$ & $y(H^0)$   &  \\
  ~\textbullet~$H^0 \to ZZ^* \to 4l$                              &                     &            & $|\cos\theta^*|$ &  \\
  ~\textbullet~$H^0 \to W^+W^- \to e^{\pm}\mu^{\mp} \nu \bar \nu$ &                     &            & $\Delta\phi(\gamma \gamma)$ &  \\
                                                                  &                     &            & $m_{34}(ll)$ & \\
                                                                  &                     &            & $y(ll)$ & \\
 & & & &\\
\hline
 & & & &\\
 Associated gauge boson and    & $g^* + g^*\to V + Q +\bar Q$   & $p_T(\gamma)$     & $p_T(V)$ & \cite{49, 50} \\
 heavy flavor production       & $g + Q\to V + Q$               & $p_T(Q)$          & $p_T(Q)$ &   \\
 ~\textbullet~$\gamma + Q$     & $q + Q\to V + q + Q$           & $m(Z/\gamma^*)$   & $y(V)$ &     \\
 ~\textbullet~$Z/\gamma^* + Q$ & $q + \bar q\to V + Q +\bar Q$  & $m_T(Z/\gamma^*)$ & $y(Q)$ &     \\
                               &                                &                   & $\Delta R(ZQ)$ & \\
                               &                                &                   & $\Delta \phi(ZQ)$ & \\
 & & & &\\
\hline
\end{tabular}
\end{center}
\caption{List of the available processes. Note that exact definitions of all kinematical variables can be found, for example,
  in corresponding references.}
\end{table}

List of processes, currently available in the \textsc{pegasus}, is presented in Table~2.
We would like to clarify some points, which are not mentioned in the Table~2:
\begin{itemize}
\item $Q$ denotes a heavy ($c$ or $b$) quark, ${\cal Q}$ denotes
$B_c$, $B_c^{(*)}$, $J/\psi$, $\psi^\prime$, $\Upsilon(3S)$, $\chi_{cJ}(1P)$
or $\chi_{bJ}(3P)$ mesons with $J = 0$, $1$ or $2$, $V$ stands for $\gamma$ or
$Z/\gamma^*$ and $l = e$ or $\mu$.
\item The standard spectroscopic notation $^{(2S+1)}L_J^{(a)}$ is used for 
intermediate Fock state of heavy quark pair $Q\bar Q$ produced 
with spin $S$, orbital angular momentum $L$, total angular momentum $J$
and color representation $a$.
\item Besides the scales, listed in the Table~2, for every process some universal scales are available, namely: 
total energy of the partonic subprocess $\sqrt{\hat s}$, also $\sqrt{\hat s/4}$ 
and so-called CCFM scale $\sqrt{\hat s+\mathbf Q_T^2}$\cite{19,30}, where $\mathbf Q_T$ is the total 
tranverse momentum of the final partons, see Fig.~3.
\item To estimate the scale uncertainties, the hard scales above can be varied 
around its default value by a factor of $2$ or $1/2$, as it often done in the pQCD calculations. 
This applies to any scale, except the CCFM scale. 
For CCFM-based gluon densities the scale 
uncertainties are evaluated by a change of the default gluon distribution 
to so-called "+" and "--" ones of the corresponding set\cite{19,30}.
\item Part of the non-logarithmic loop corrections to effective $g^* + g^* \to H^0$ vertex
can be absorbed in the special $K$-factor\cite{51} and optionally implemented into the calculations. 
As default choice, this $K$-factor is switched off.
\item According to experimental setup, an isolation criterion is applied for prompt photon production. 
This criterion is the following: a photon is isolated if the amount of hadronic transverse energy
$E_T^{\rm had}$ deposited inside a cone with aperture $R$ centered around
the photon direction in the pseudo-rapidity and azimuthal angle plane, is smaller 
than some value $E_T^{\rm max}$ ("cone isolation"). 
The isolation requirement significantly (up to $\sim 10$\% of the
visible cross section) reduces contribution from the 
so-called photon fragmentation mechanisms, not implemented into the \textsc{pegasus}.
The isolation criterion and 
additional conditions which preserve our calculations from divergences 
have been specially discussed, for example, in\cite{52} (see also references therein).
The default values $R = 0.4$ and $E_T^{\rm max} = 4$~GeV
can be changed optionally.
\item At present, quark-initiated subprocesses can be calculated 
only within collinear QCD factorization and no TMD quark densities are 
implemented into the current version of \textsc{pegasus}. 
The QCD Compton subprocess is available only, if collinear 
factorization has been chosen, since in the $k_T$-factorization approach 
its contribution is taken into account by the off-shell gluon-gluon fusion (see\cite{49,50} for more 
details).
\end{itemize}

\subsection{Quarkonium final states} \indent

Quarkonium production processes need additional explanations as they
contain an important extra step: the formation of bound states.

The process starts with the production of a heavy quark-antiquark pair $Q \bar Q$ 
in a hard parton interaction. The produced quark pair may be either color
singlet or color octet. If the $Q \bar Q$ state is color singlet, it can
immediately convert into a meson with appropriate quantum numbers. Then 
the formation probability is determined by a single parameter, the radial
wave function at the origin of the coordinate space, $|{\cal R}_S(0)|^2$ or 
$|{\cal R}'_P(0)|^2$\cite{53,54,55,56,57}.
The values of these functions can be extracted from the measured decay
widths or calculated within potential models.

The situation with color-octet states is more complicated as the 
transition to a color-singlet physical hadron requires the emission 
of extra (soft) gluons. Then, for every final state hadron $h$ and 
for every intermediate $Q \bar Q$ state $n = \,^{2S+1}L_J$ 
listed in Table~2 there is 
a specific phenomenological long-distance matrix element (LDME)\cite{58,59,60} 
responsible for such a transition
$\langle {\cal O}^{h} \left[n\right]\rangle$.
Eventually, the production cross section for a hadron $h$ in $pp$ collisions 
is given by a sum over all possible singlet and octet $Q\bar Q$ states:
\begin{equation}
  \sigma(pp\to h + X) = \sum_{n} \sigma\left(pp\to Q\bar Q\left[n\right]\right) \langle {\cal O}^{h}\left[n\right]\rangle,
\end{equation}
\noindent
where $\sigma\left(pp\to Q\bar Q\left[n\right]\right)$ is the partial partonic cross section~(1)
for a $Q\bar Q$ state $n$.
The different states $n$ are selected by introducing 
the proper projection operators in the hard scattering amplitude.  
The correspondence between the color singlet and color octet wave functions and respective LDMEs is given by the 
$\langle {\cal O}^{h}\left[\,^{2S + 1}L_{J}\right]\rangle = 2N_c(2J+1)|{\cal R}_{S}(0)|^2/{4\pi}$
and $\langle {\cal O}^{h}\left[\,^{2S + 1}L_{J}\right]\rangle = 6N_c(2J+1)|{\cal R}_{P}^\prime(0)|^2/{4\pi}$, respectively.

As default choice to describe the spin structure of relevant transition amplitudes, \textsc{pegasus} uses 
the model\cite{61}, where the NRQCD emission of soft gluons is
considered in terms of classical multipole radiation theory. The multipole
expansion is dominated by (chromo)electric dipole transitions $E1$.
According to\cite{61}, only a single $E1$ transition is needed to transform a $P$-wave state 
into an $S$-wave state and the structure of the respective 
${^3P_J^{(8)}}\to {^3S_1^{(1)}} + g$ amplitudes is taken the same as for radiative 
decays of $\chi_{cJ}$ mesons\cite{62,63}:
\begin{equation}
  {\cal A}(\,^3P_0^{(8)} \to \, ^3S_1^{(1)} + g) \sim k_\mu \, p^\mu \, \epsilon_\nu (l) \epsilon^\nu(k),\label{3p0}
\end{equation}
\begin{equation}
  {\cal A}(\,^3P_1^{(8)} \to \, ^3S_1^{(1)} + g) \sim e^{\mu \nu \alpha \beta} k_\mu \, \epsilon_\nu(p) \, \epsilon_\alpha (l) \epsilon_\beta(k),\label{3p1}
\end{equation}
\begin{equation}
  {\cal A}(\,^3P_2^{(8)} \to \, ^3S_1^{(1)} + g) \sim p^\mu \, \epsilon^{\alpha \beta}(p) \, \epsilon_\alpha (l) \left[ k_\mu \epsilon_\beta(k) - k_\beta \epsilon_\mu(k) \right], \label{3p2}
\end{equation}
\noindent
where $p$, $k$ and $l = p - k$ are the four-momenta of the color-octet $P$-wave state, 
emitted gluon and produced color-singlet $S$-wave state, 
$\epsilon^\mu(k)$, $\epsilon^\mu(l)$, $\epsilon^\mu(p)$ and $\epsilon^{\mu\nu}(p)$ are 
the polarization vectors (tensor) of respective particles and $e^{\mu \nu \alpha \beta}$
is the fully antisymmetric Levi-Civita tensor.
The transformation of an $S$-wave state into another $S$-wave state 
(such as $J/\psi$ or $\psi^\prime$ meson) is treated as two successive $E1$ transitions
$^3S_1^{(8)}\to \,^3P_J^{(8)}+g$, $^3P_J^{(8)}\to \,^3S_1^{(1)} + g$
proceeding via either of the three intermediate states:
$^3P_0^{(8)}$, $^3P_1^{(8)}$, or $^3P_2^{(8)}$. For each of the two transitions 
the same effective coupling vertices (23) --- (25) are exploited.
In the case of $J/\psi$ or $\Upsilon(3S)$ production,
the user can optionally include feed-down from the decays of upper quarkonium
states ($\psi^\prime$, $\chi_{cJ}(1P)$ or $\chi_{bJ}(3P)$, respectively)
in addition to the direct production channels.

The absolute normalization of the transition amplitudes is not calculable
within the theory; these numbers are taken as free adjustable parameters.
However, the values of the LDMEs for the different partial contributions 
are not completely independent but are connected by the heavy quark spin 
symmetry (HQSS) relations\cite{60}. 
All the HQSS relations between LDMEs are implemented in the 
\textsc{pegasus} default 
setting, taken from\cite{46,47}.


\subsection{Strong coupling and masses of particles} \indent

In \textsc{pegasus}, the strong coupling $\alpha_s$
can be calculated in one-loop or two-loop approximation
with respect to the number of active flavors $N_f$
and $\Lambda_{\rm QCD}$.
The choice of $N_f$ and $\Lambda_{\rm QCD}$ is done automatically
(according to the Table~1)
with the choice of the TMD and/or conventional parton 
density in a proton.
There is no possibility to change it manually
since this setup is essential for determination of corresponding
parton distributions.

The masses of all particles (quarks, gauge bosons, heavy quarkonia etc),
their branching ratios and decay widths are fixed according to 
Particle Data Group (PDG)\cite{64}.
Any of these parameters can be easily changed using the convenient 
built-in \textsc{particle data} tool (see Fig.~6),
which is available via main menu ({\sl Edit $\to$ Settings $\to$ Particle data})
or popup menu or appropriate button on the button panel.

\subsection{Generation of Les Houches Event file} \indent

\textsc{pegasus} is supplied with a tool to construct event files in the Les Houches Event format\cite{29}.
This is a widely accepted format to present events, which is compatible with the majority of modern general 
purpose Monte-Carlo generators.

The Les Houche Event (LHE) file, generated by the \textsc{pegasus}, consists of two main blocks: the first one contains the
information about the number of the recorded events, the PDFs used, the colliding hadrons and their energies.
Also the total cross section is shown. The second block represents a list of events, including the data on the interacting partons,
their 4-momenta and color structure of the event.
We mention the basic features of the LHE file, generated by the \textsc{pegasus}:
\begin{itemize}
\item The generated events could be weighted or unweighted. In first case, 
the sum of all the weights is the total cross section of the subprocess.
\item Polarization information is not preserved.
\item A parton carries a tag according to the standard PDG numbering scheme\cite{64}.
\item Conventional (collinear) parton densities in a proton are numbered according to the \textsc{lhapdf} scheme\cite{65} 
  while TMD parton distributions are numbered according to the \textsc{tmdlib} package\cite{39}.
\end{itemize}

The produced LHE file can then be processed with an external program to introduce some peculiar event selection,
to include parton showers, to hadronize the final particles, etc. It is found
to be compatible with such Monte Carlo generators as \textsc{pythia}\cite{6} and \textsc{cascade}\cite{9}.

\section{Program components} \indent

\subsection{Random number generator} \indent

Since all the internal variables in \textsc{pegasus} are declared as double precision ones,
double precision random numbers have to be generated in the Monte-Carlo simulations.
The random number generator \textsc{ranlux}\cite{66} is well suited for these purposes.
It has a long period, solid theoretical foundations and is commonly used in computational physics.
This random number generator is implemented into the \textsc{pegasus}.

\subsection{Phase space integration and event generation} \indent

The multidimensional phase space integration (1) is performed
with the Monte-Carlo technique and is incorporated with the routine \textsc{vegas}\cite{12}.
The routine \textsc{vegas} allows up to ten integration variables, that 
is enough for subprocesses considered in \textsc{pegasus}.

The \textsc{vegas} algorithm is based on a method for reducing statistical errors
by using a known probability distribution function to concentrate the 
search in those areas of the integrand that make the greatest contribution 
to the final integral. 

The algorithm is realised through a large number of random sample points
distributed over a $d$-dimensional rectangular volume. The whole volume
is divided into $d$-dimensional rectangular cells (by default, 50 divisions
along each axis). The probability for a point to drop into a given cell is
determined by so called sampling distribution, which is adjusted to
the integrand function. The sampling distribution approximates the 
exact distribution by making a number of passes (iterations) over 
the integration region while histogramming the integrand function 
$d\sigma$ given by the expession (1). 
Each iteration is used to define a sampling distribution for the next
iteration. To improve the convergence in the region of high $p_T$, the
user can optionally modify the integrand function from 
$d\sigma$ to $(1 + p_T^4)d\sigma$, see Fig.~1. The optimization of the sample grid
is made automatically and needs no care from the user.

Each sample point generated by \textsc{vegas} represents an event in the $n$-particle
phase space with the coordinates of the sample point responding to the 
values of the physical integration variables ($p_{iT}$, $y_i$, $\phi_i$). 
The weight attributed to that event is given by the
product of the integrand function $d\sigma$ and the $d$-volume of the 
sampling cell containing that point, $dw = d\sigma dV_{\rm cell}$.
Unweighting algorithm for the generated events is provided also.

The typical time needed to generate one event, of course,
depends on the requested subprocess; but, in general, 
is similar to time needed by other Monte-Carlo event
generators like \textsc{cascade} or \textsc{pythia}.
The output events can be plotted on a histogram (using built-in tool \textsc{pegasus plotter} ) 
or stored for further use in the form of an Les Houches Event file.

\subsection{PEGASUS Plotter} \indent

\textsc{pegasus} is supplied with a built-in tool \textsc{pegasus plotter}, 
allowing one to depict easily the produced differential cross sections and immediately 
compare them with experimental data. As a default setting, 
the accumulated results for 
requested observables during the calculation are shown in \textsc{pegasus plotter}.
However, it is a quite independent tool and can be used 
apart from any calculations made within the \textsc{pegasus}. 
The program is very simple and intuitive.
Let us briefly describe main features of the tool.

As one calls the \textsc{pegasus plotter} from the main 
menu of the \textsc{pegasus} (using the {\sl Tool $\to$ Plotter} option, or via popup menu, or by pressing the corresponding 
button on button panel) an empty sheet is created. 
The following objects, stored in a plain data files, could be added to the sheet (by choosing {\sl Edit $\to$ Add} 
option in the main menu or popup menu
available with a right mouse button click on the sheet):
\begin{itemize}
\item {\it Curve}. The data file should consist of two columns, corresponding to the rows of $x$ and $y$ values.
\item {\it Histogram}. The data file is the same, as for {\it Curve}. However, every $y$ value should be mentioned 
  twice, for the both borders of the corresponding bin. 
\item {\it Filled area}. A three-columns data file should contain for every $x$ value lower and upper values of $y$.
\item {\it Text label}. An arbitrary text note inside the plot sheet. 
Greek letters are available through the syntax \texttt{\{/Symbol letter\}}, where the \texttt{letter} is just
the name of the greek symbol (for instance, \texttt{alpha}, \texttt{sigma} or other). Several capital greek symbols
are available, namely, $\Upsilon$ (\{/Symbol \texttt{Upsilon}\}), $\Psi$ (\{/Symbol \texttt{Psi}\}) and
$\Delta$ (\{/Symbol \texttt{Delta}\}).
Subscripts and superscripts are available with a latex-like syntax \texttt{\_\{subscript\}}, \texttt{\^{ }\{superscript\}}.
\item {\it Experimental data}. The data file should be in \textsc{gnuplot}-compatible format with 6 
columns, corresponding to $x$, $y$, lower and upper $x$ values and lower and upper $y$ values. Alternatively, 
the data files in standard \texttt{*.yoda} or \texttt{*.csv} format (available from HepData repository\cite{28})
can be uploaded.
\end{itemize} 

As an object is added on the sheet, it can be selected with a left mouse button click and modified 
according to user own wishes either with a double click or with 
choosing option {\sl Edit$\to$ Plottable} in the \textsc{pegasus plotter} main menu or via popup menu. 
Then the text label in the legend and 
appearance of the selected object (for example, color, font, size etc) can be changed. 
If selected object is a {\it Histogram}, the fiducial cross section (integral with respect to the $x$ variable)
is shown in the status bar.
One can also set a factor to scale the depicted cross sections using {\sl Edit $\to$ Multiply by a factor}
option in main menu or popup menu.

The default axes setting can be changed 
from the main menu ({\sl Edit $\to$ Axes} option) or by double clicking an axis. 
Besides the font, alignment and other setting one can also set the axes to be linear or 
logarithmic. 
From the main menu ({\sl Options $\to$ Plot size}) or popup menu
one can also adjust the size of the graph in pixels.

The plot can be saved for the future editing
via main menu options {\sl File $\to$ Save} or {\sl File $\to$ Save As} or via popup menu
in the internal format (\texttt{*.pplot}). The export  
to a \textsc{gnuplot} script is possible via main menu {\sl Export $\to$ Plot to Gnuplot script} option or via 
popup menu.
The figure can be also printed out or saved in \texttt{*.png}, \texttt{*.jpg} or \texttt{*.bmp} format.
Finally, samples for all plotted curves, histograms or data point sets (or for only selected ones) 
can be transfered (using {\sl Options $\to$ Export}) to a plain data file (which is compatible, for example, with \textsc{gnuplot})
for future usage in other programs.

Some typical snapshots of the \textsc{pegasus plotter} are presented on Fig.~7.

\section{Installation and running} \indent

\textsc{pegasus} can be downloaded from \url{https://theory.sinp.msu.ru/dokuwiki/doku.php/pegasus/download} 
as a precompiled executive
file for Linux machines. No special installation procedure is needed.
The data files containing the necessary TMD parton densities in a proton 
(and conventional PDFs as well) are located
inside the \textsc{pegasus} home directory (folder \texttt{data}).
If there are some missing data files in \texttt{data} folder, \textsc{pegasus} will inform user 
about that (see Fig.~?). In this case, no calculation is allowed.
Location of \texttt{data} folder could be easily changed via main menu option {\sl Edit $\to$ Settings $\to$ Path to data folder}. 

For Linux machines, the executive file can be just run from a terminal as \texttt{./PEGASUS}.
The program demands the \textsc{qwt} library (version 6.1.3)\cite{11}, so the library file \texttt{libqwt.so.6} 
should be inside the \textsc{pegasus} home directory. Otherwise, the path to
this file should be specified with \texttt{export LD\_LIBRARY\_PATH=/path/to/library}.

The program was tested on ROSA Linux R8.1, ROSA Linux R11 and Ubuntu 16.04.

\section*{Acknowledgements} \indent

We would like to thank all our colleages and friends, who supported us during work on this program.
We thank Victor Savrin and Edward Boos for their encouraging interest and creating comfortable and friendly 
atmosphere in SINP MSU. Special thanks to Elena Boos. Without her the way of this program to the release would be
much longer.

A part of the work was done during author's visits in DESY (Hamburg, Germany). We are grateful the DESY 
Directorate for the support in the framework of the Cooperation Agreement between MSU and DESY. 
We thank specially Hannes Jung and DESY CMS QCD group for warm hospitability and valuable feedback.

We are also grateful to Maria Mikova, Natalia Ovechkina and Anastasia Zotova for their support and help 
for the design of the program.

We are especially grateful to Nikolai Zotov, who guided and supervised our first steps in High Energy Physics, 
who encouraged our progress in $k_T$-factorization and whose enthusiasm consolidated our group.
Nikolai Zotov passed away on January 2016. Our work is dedicated to his memory.

M.A.M. was supported in part by a grant of the foundation for the advancement
of theoretical physics and mathematics ”Basis” 17-14-455-1.

\newpage

\begin{figure}
\begin{center}
\includegraphics[width=13cm]{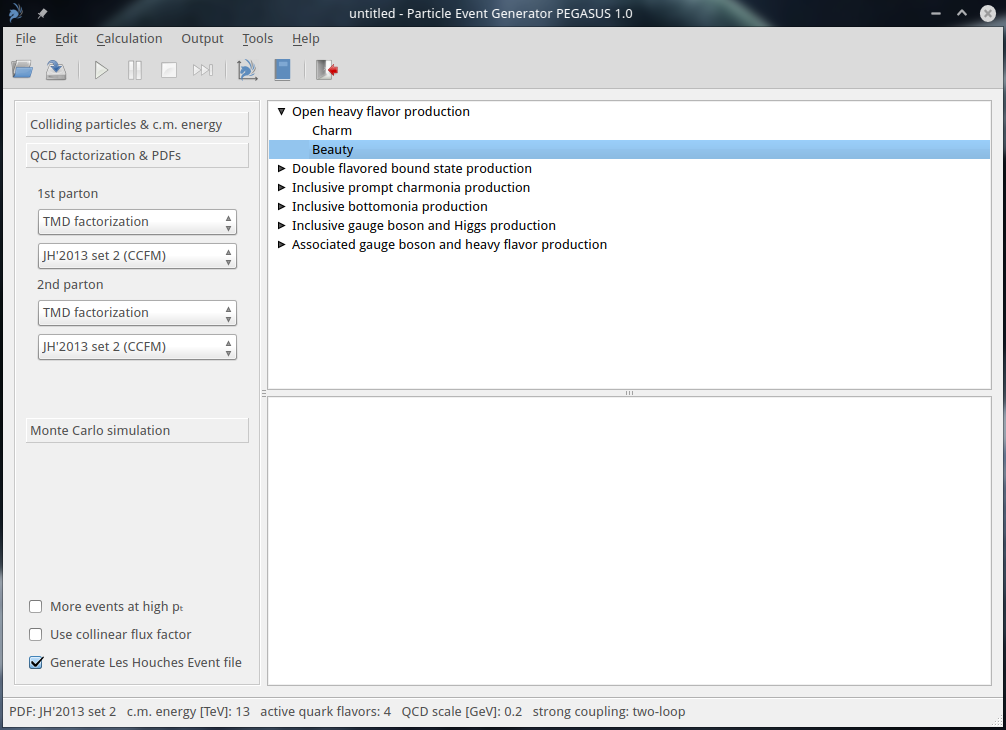}
\caption{\textsc{pegasus} main window. Here one can select factorization scheme (TMD or collinear one)
for each of the colliding particles, choose corresponding parton density function and set the parameters, important 
for further Monte-Carlo simulation: number of iterations and number of simulated events per iteration.}
\label{fig1}
\end{center}
\end{figure}

\newpage

\begin{figure}
\begin{center}
\includegraphics[width=13cm]{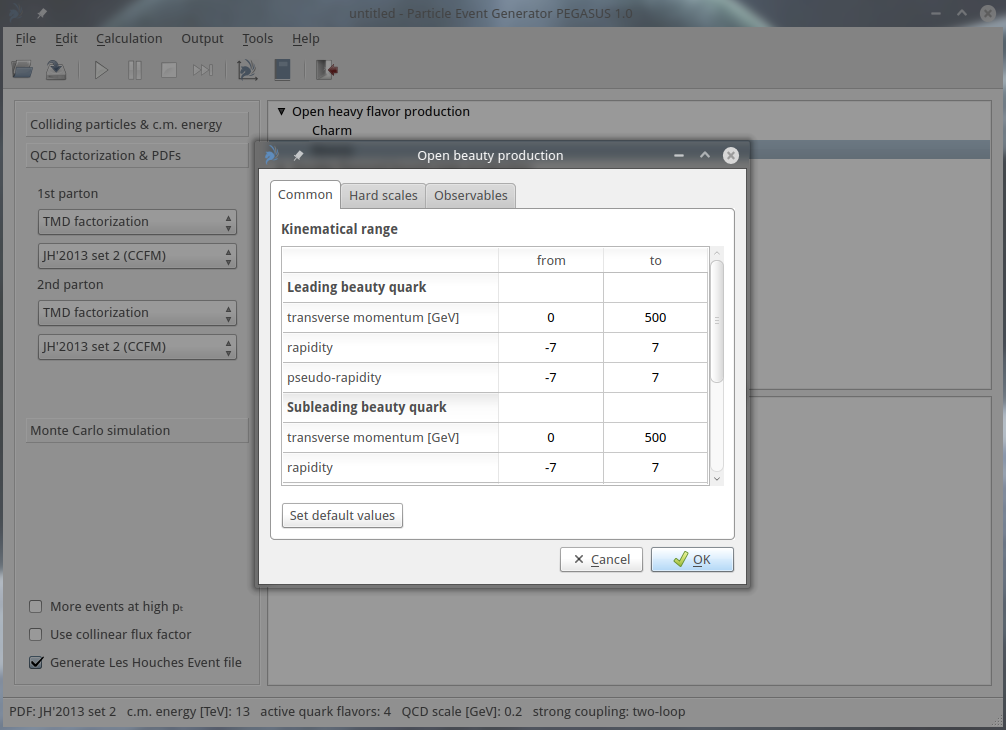}
\caption{One can optionally correct the default kinematical 
restrictions, list of requested observables and corresponding binnings for any processes (part 1).}
\label{fig2}
\end{center}
\end{figure}

\newpage

\begin{figure}
\begin{center}
\includegraphics[width=13cm]{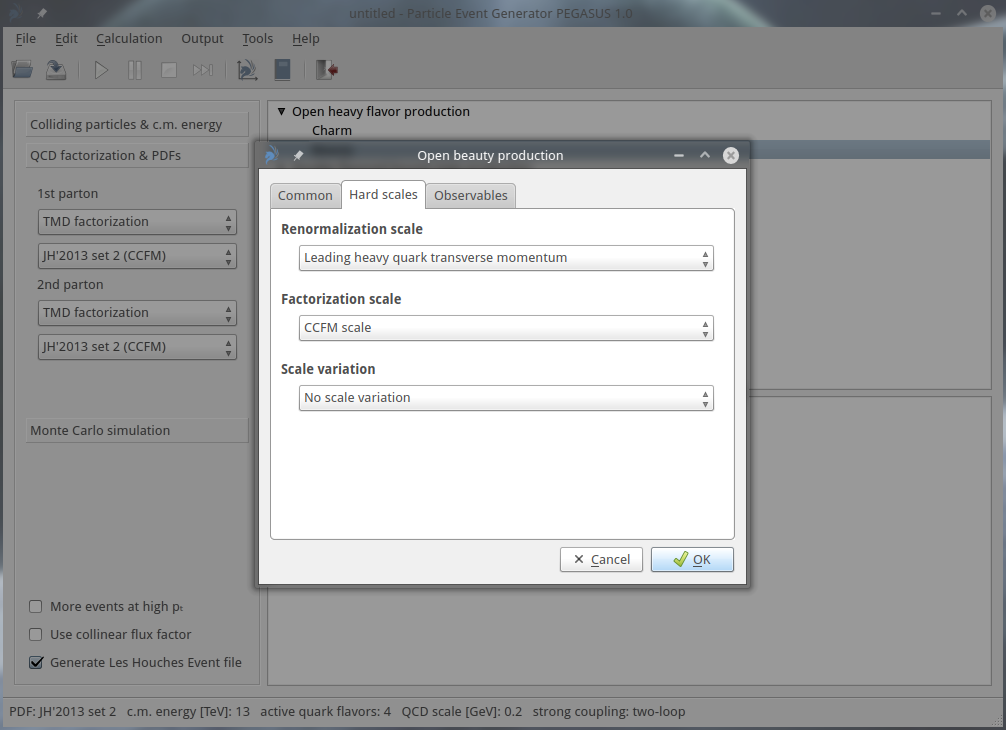}
\caption{One can optionally correct the default kinematical 
restrictions, list of requested observables and corresponding binnings for any processes (part 2).}
\label{fig3}
\end{center}
\end{figure}

\newpage

\begin{figure}
\begin{center}
\includegraphics[width=13cm]{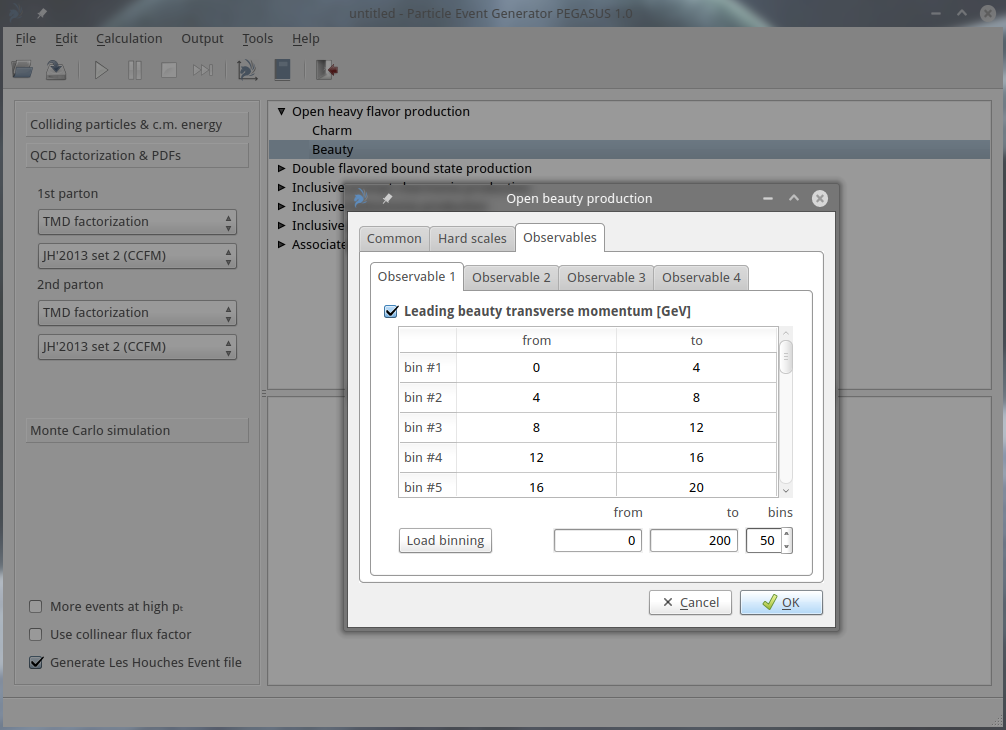}
\caption{One can optionally correct the default kinematical 
restrictions, list of requested observables and corresponding binnings for any processes (part 3).}
\label{fig4}
\end{center}
\end{figure}

\newpage

\begin{figure}
\begin{center}
\includegraphics[width=13cm]{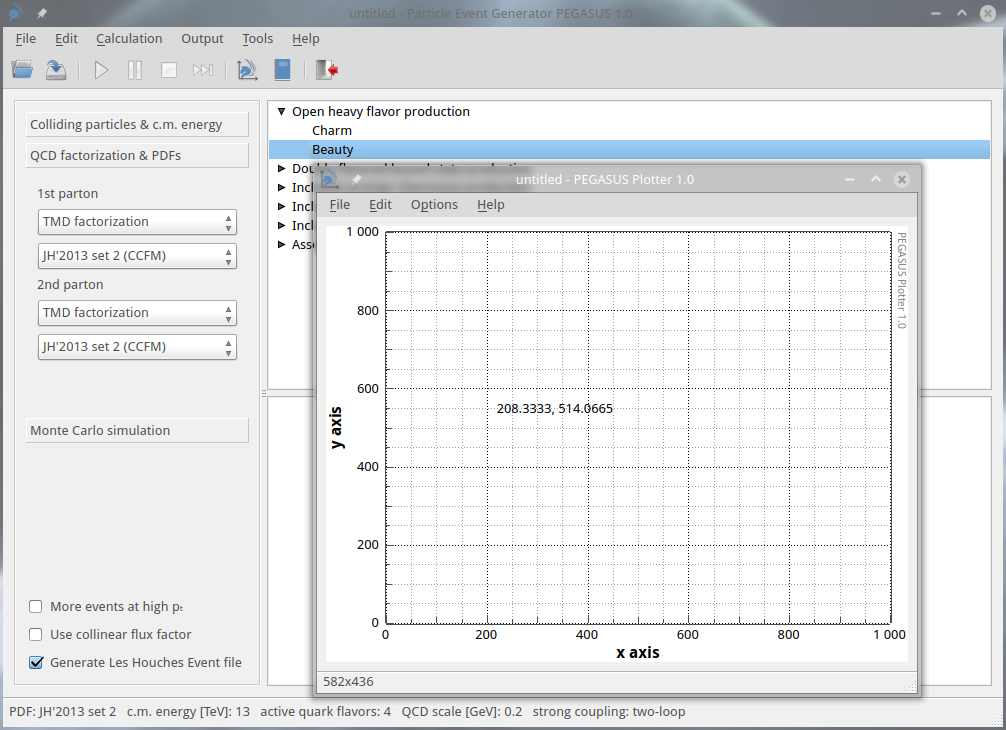}
\caption{\textsc{pegasus plotter} can be launched from \textsc{pegasus} main window.}
\label{fig5}
\end{center}
\end{figure}

\newpage

\begin{figure}
\begin{center}
\includegraphics[width=13cm]{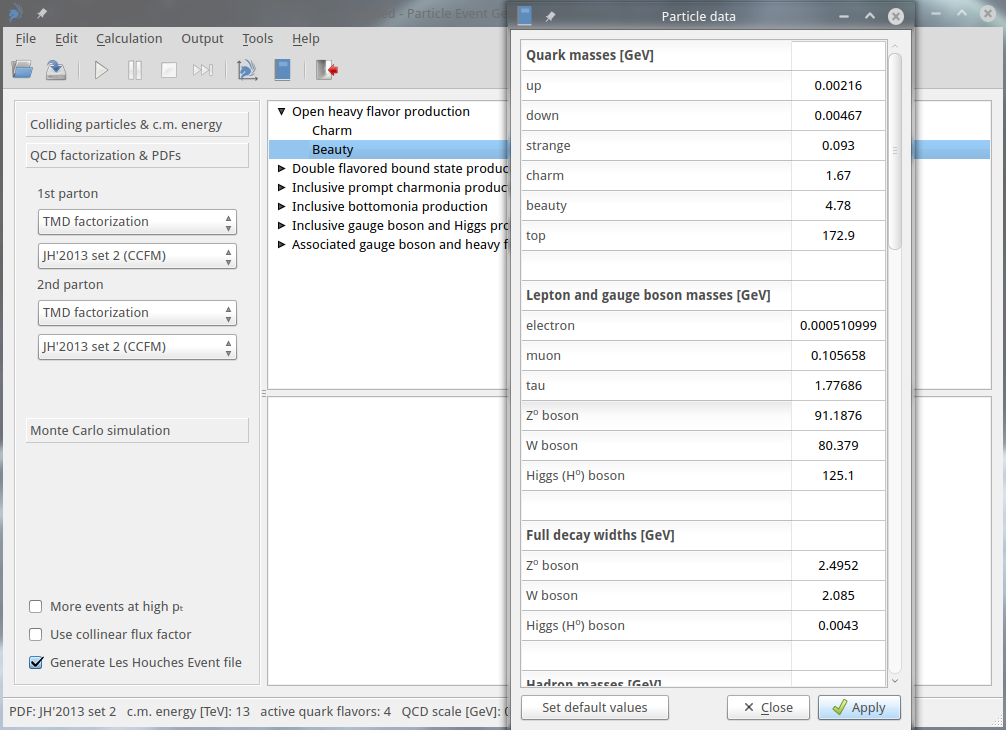}
\caption{\textsc{Particle data} tool can be launched from \textsc{pegasus} main window.}
\label{fig6}
\end{center}
\end{figure}

\newpage

\begin{figure}
\begin{center}
\includegraphics[width=13cm]{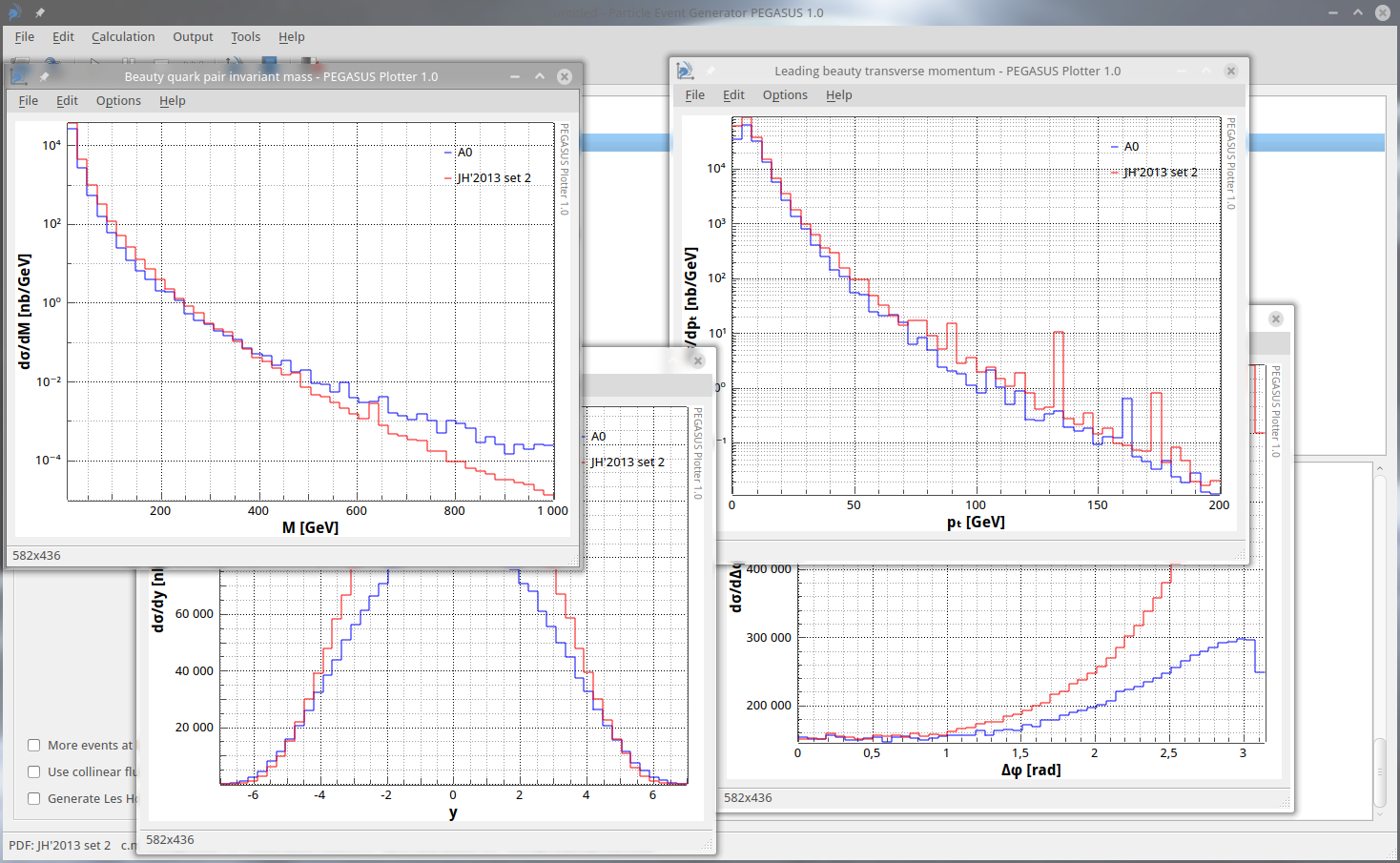}
\caption{Results of the calculations presented by \textsc{pegasus plotter}.}
\label{fig7}
\end{center}
\end{figure}

\newpage

\begin{figure}
\begin{center}
\includegraphics[width=13cm]{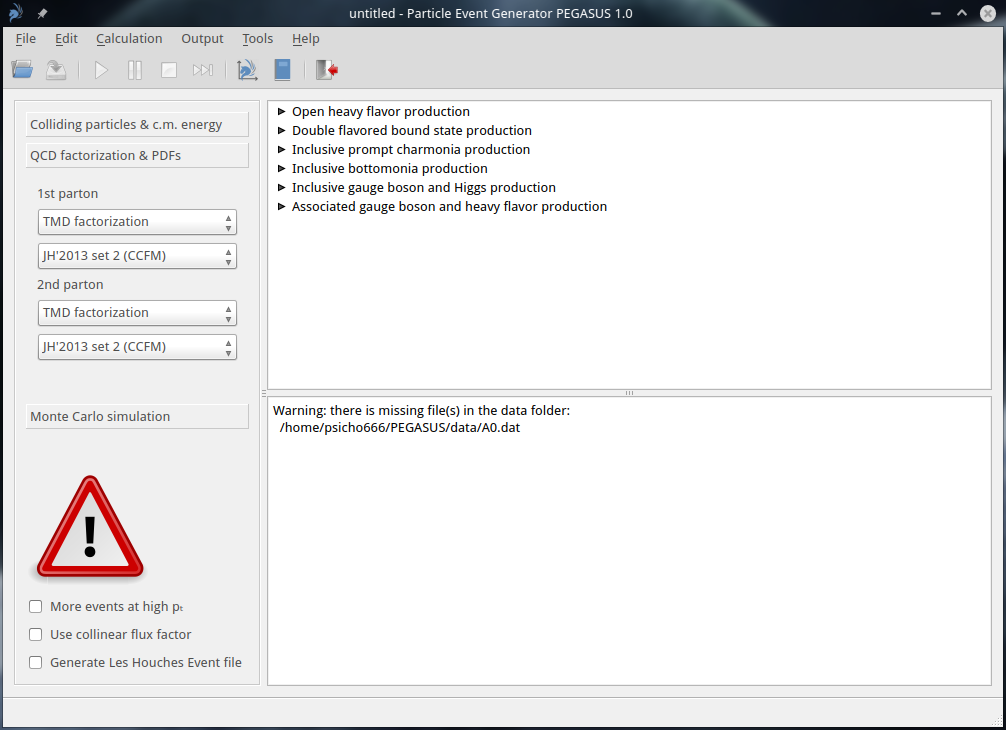}
\caption{If there are some missing data files in \texttt{data} folder, \textsc{pegasus} informs user 
about that.}
\label{fig8}
\end{center}
\end{figure}

\end{document}